# First direct determination of the Boltzmann constant by an optical method

C. Daussy, M. Guinet, A. Amy-Klein, K. Djerroud, Y. Hermier[1], S. Briaudeau[1]
Ch.J. Bordé, and C. Chardonnet
Laboratoire de Physique des Lasers, UMR CNRS 7538,
Institut Galilée, Université Paris 13, 99, ave J.-B. Clément – 93430 Villetaneuse – France
[1] Permanent address : Institut national de Métrologie LNE-INM – CNAM – La Plaine Saint-Denis

Abstract

We have recorded the Doppler profile of a well-isolated rovibrational line in the $\nu_2$ band of $^{14}NH_3$. Ammonia gas was placed in an absorption cell thermalized by a water-ice bath. By extrapolating to zero pressure, we have deduced the Doppler width which gives a first measurement of the Boltzmann constant, $k_B$, by laser spectroscopy. A relative uncertainty of $2\times10^{-4}$ has been obtained. The present determination should be significantly improved in the near future and contribute to a new definition of the kelvin.

The tremendous progress in high precision measurements during recent decades will lead unavoidably to a complete renewal of fundamental metrology. There is a strong tendency to relate the base units to fundamental constants [1]. As an example, this has been done in 1983 by fixing the velocity of light, $c$ and thus defining the length unit from the time unit, because the second is the fundamental unit which is realized, by far, with the best accuracy. The unit of temperature could follow the same line. Up to now, the kelvin is defined by the temperature (273.16 K) of the triple point of water (TPW) which implies a particular property of macroscopic matter. Instead, the temperature of a sample has a microscopic interpretation and can be related through the Boltzmann constant to the mean energy, $E$, per particle and per degree of freedom according to the well-known expression: $E = 1/2\, k_B T$. This energy may itself be related to a frequency via Planck constant. This paper presents a first accurate experiment which gives a direct measurement of such a frequency, in a gas at a well-defined temperature. Fixing the value of $k_B$ would connect temperature and time units. But, before fixing the value of the Boltzmann constant it is necessary to verify precisely the consistency of the value of $k_B$ in the present context. The accepted value in the CODATA [2], $k_B=1{,}380\ 6505\ (24) \times 10^{-23}$ J K$^{-1}$, is derived from the value of the ideal gas constant, $R$, and the Avogadro constant $N_A$, by the relation: $k_B=R/N_A$. The relative uncertainty of $k_B$ is $1.8\times10^{-6}$ and should come mostly from that for $R$ because the uncertainty of $N_A$ is $1.7\times10^{-7}$ [2]. But, there is presently an inconsistency at the level of $10^{-6}$ between the values of $N_A$ derived from the Si sphere and from the watt balance experiment [3].

Very few experiments lead to an accurate determination of $k_B$ or $R$ [4]. Up to now, the accepted value of $R$ comes from a single experiment by Moldover et al. [5] performed before 1988 by acoustic gas thermometry.

An alternative and indirect measurement of the Boltzmann constant was proposed along an approach based on the virial expansion of the Clausius-Mossotti equation [6]. This relates the permittivity of helium, $\varepsilon$, to its molar polarisability, $A_\varepsilon$, which implies a QED calculation.

Here, we propose a direct determination of the Boltzmann constant by laser spectroscopy [1]. The principle consists in recording the linear absorption in vapour phase and measuring the Doppler width of an atomic or molecular line in a cell at the thermodynamic equilibrium. In the Doppler limit, the line shape is a Gaussian (for an optically thin medium) and $k_B T$ is given by:

$$k_B T = \frac{mc^2}{2}\left(\frac{\Delta_D}{\nu}\right)^2$$

$\Delta_D$ is the e-fold half-width, $\nu$ is the frequency of the molecular line and $m$ is the molecular mass. $\Delta_D$ and $\nu$ are determined experimentally. The probed atoms or molecules belong to a single quantum level of a well-defined isotopic species, which avoids uncertainties coming from macroscopic quantities as it is the case in the schemes mentioned above. Strictly speaking, we are sensitive to the temperature of one translational degree of freedom of a subset of molecules. Since the temperature is measured on the walls of the cell containing the gas, the determination of $k_B$ using different rovibrational lines can lead to a verification of the equipartition of energy principle. This experimental situation benefits from an analysis considerably more straightforward than other methods, as we will see.

Ion mass ratios can now be measured in Penning traps with $10^{-9}$-$10^{-10}$ accuracies [7] and binding energies for a molecule can be easily calculated to keep the accuracy of the molecular mass expressed in atomic units at the same level. However, because of the definition of the kilogram, its absolute value implies the Avogadro constant, $N_A$, as in the previous experiments. Since atom interferometry yields a direct determination of the quantity $h/mc^2$ [8], the present experiment yields directly the ratio, $k_B/h$.

**Experiment**

The experiment consists simply in recording an absorption spectrum. The selected line for these first experiments is the $\nu_2$ asQ(6,3) rovibrational line of the ammonia molecule $^{14}NH_3$ at the frequency $\nu = 28\ 953\ 694$ MHz. The choice of the molecule is governed by two main reasons: –a strong absorption band in the spectral region of 8-12 µm of the ultra-stable spectrometer that we have developed over several years; –a well-isolated Doppler line to avoid any deformation of the line shape due to neighbouring lines. A light molecule, such as ammonia, fulfils these conditions with the advantage of a large absorption line. Clearly, the conditions for such an experiment can be obtained with a wide set of molecules or even atoms in various spectral regions and also over a large range of temperatures. In our experiment, the gas pressure varies from 0.1 to 10 Pa, which is 4 orders of magnitude smaller than the pressure of the acoustic gas thermometry or permittivity experiments. This is a major advantage for any zero-pressure extrapolation. The 30 cm-long absorption cell is placed in a large thermostat filled with an ice-water mixture, fixing the temperature at 273.15 K. Several temperature probes based on 100 Ω platinum resistors with a 0.39 Ω K$^{-1}$ dependency are firmly attached to the cell to follow the temperature changes or gradients. These probes were calibrated against a TPW. A conservative uncertainty of 20 mK ($7\times10^{-5}$) was guaranteed in this experiment. This can be easily improved in the future. The laser source is based on a $CO_2$ laser stabilized on an absorption line of $OsO_4$. The laser frequency control is a key point of the experiment: the spectral purity is better than 10 Hz and the Allan Variance @ 100 Hz can reach 0.1 Hz [9]. Recently, we have measured the absolute frequency of our laser source against the Cs fountain of SYRTE (Paris Observatory) through an optical fibre link connecting our two laboratories. The uncertainty of 1 Hz ($3\times10^{-14}$) [10] shows that the relative precision of the frequency axis of the recorded Doppler profile (width of about 100 MHz) can reach $10^{-8}$. This is 2 orders of magnitude better than the actual accuracy of the Boltzmann constant and will not be a limitation. The tunability of the laser source is obtained with an 8-18 GHz electro-optic modulator (EOM) [11] which generates two weak sidebands (SB) ($10^{-4}$-$10^{-3}$ efficiency). The sideband, red-detuned by 13 GHz from the R(6) $CO_2$ laser carrier, is put into resonance with the desired molecular transition. It is tuned to record the full Doppler profile by detecting the light transmitted by the cell on a HgCdTe photo-detector. The main difficulty in this experiment is our capability to record a signal which reflects perfectly the Doppler profile and to suppress any parasitic optical signals. For this, several precautions were taken step by step and the present set-up is displayed on Fig.1. A polarizer blocks 99.5% of the laser carrier and transmits the 2 cross-polarized sidebands. Then, a Fabry-Perot cavity (FPC) with a free spectral range of 1 GHz and a finesse of 150 is used to filter drastically the residual carrier and the undesired sideband. For that purpose, the sidebands via the EOM are frequency modulated at f=8 kHz with a depth of 38 kHz to lock the FPC on the useful sideband. However, when the EOM is tuned around the molecular line, resonance conditions can be reached accidentally by different transverse modes of the carrier or the other sideband. This problem may be strongly reduced by a careful choice of the laser frequency, which can be adjusted via a second EOM used for the laser stabilization (not represented on Fig. 1). To eliminate even more the laser carrier residuals, a 30% amplitude modulation at $f_1$=17 kHz is applied to the sidebands only via the 8-18 GHz EOM. With this successive filtering, the contrast of the selected sideband against carrier and unwanted sideband is better than 1000. The next main concern is the stability of the laser intensity during the tuning range. In fact, the response of our EOM can vary by 20% over the 300 MHz range used for recording the molecular spectrum. To circumvent this unacceptable problem we split the laser beam into two channels: the reference beam (A) is frequency shifted with an acousto-optic modulator (AOM) of 80 MHz and amplitude modulated at $f_2$=1.7 kHz; the probe beam (B) which traverses the absorption cell is also frequency shifted by an AOM of 40 MHz and amplitude modulated at $f_3$=2.2 kHz. The path lengths are adjusted to be equal and the two beams are recombined to be directed to the same photo-detector. The optical set up permits, first, demodulation of the combined signals at $f_1$ which eliminates any contribution of the laser carrier. Then, each signal is recovered by demodulations at $f_2$ and $f_3$. The reference signal reflects the sideband intensity and is used to apply a correction to the amplitude of the EOM signal to maintain it constant during the frequency scan. Thus, the signal detected at $f_3$ reflects exactly the absorption signal of the molecular gas recorded with a constant incident laser power. Starting from a laser carrier of 1.5 W, the available power in the cell can be chosen between 0.1 µW to 3 µW for a beam diameter of 11 mm. The Doppler profile of about 100 MHz width is recorded over 250 MHz by steps of 500 kHz with a time constant of 20 ms. However, the important frequency step which is also applied to the filtering cavity by the way of a servo loop imposes a time delay of about 200 ms between two points to prevent extra noise. The time for recording a spectrum is thus about 110 sec, dominated by the waiting time. Fig. 2 displays a series of spectra recorded at different pressures. The vertical axis is the relative transmission normalized by the incident laser power.

**Methodology of the line shape analysis**

The theoretical description of this experiment corresponds to the well-established situation of the linear absorption of a sample in a cell at thermal and pressure equilibrium. However, an accurate determination of the line width requires a very accurate description of the line shape. In the Doppler limit, the line shape is a Gaussian as mentioned above. Experimentally, we try to work as closely as possible to this regime. In practice, several causes of broadening must be taken into account. In the first place, the pressure broadening and the natural width are responsible for the homogeneous width, $\gamma_{hom}$. In this case the line shape is a Voigt profile that is a convolution of a Gaussian and a Lorentzian. In linear absorption spectroscopy and isotropic velocity distribution, transit effects are fully included in the Doppler profile. Two other effects must be considered: the unresolved hyperfine structure of the transition and the modulations applied for experimental reasons which broaden the laser spectrum. In a first approach, these two effects act in the same way: in the first case, the line shape is exactly the sum of the individual hyperfine components and in the second case, the frequency comb of the laser spectrum due to the frequency modulation, $f_1$, (which is the main effect, here) generates absorption signals which reflect the laser spectrum (positions and relative intensities). We have checked that our experiments were performed in the low field limit, far from saturation. In the final line shape, one can take into account exactly all of these effects and it is possible to represent the signal by the theoretical line shape. However, since experimentally we are very close to the Doppler limit, the actual line shape cannot easily be distinguished from a broadened Gaussian. We calculated numerically the relative broadening due to the various causes mentioned above: For the homogeneous broadening, it is $0.484\, \gamma_{hom}/\Delta_D$; hyperfine structure $\Delta_{hyp}$ causes a broadening of $0.254 (\Delta_{hyp}/\Delta_D)^2$, and laser frequency modulation depth, a broadening of $(depth/\Delta_D)^2$. $\Delta_{hyp}$ is the total width of the structure and the coefficient 0.254 corresponds to the less favourable case of a doublet of equal amplitude. For the present line, the hyperfine structure of 12 components of the order of 150 kHz can be calculated or measured by saturation spectroscopy [12]. Its contribution to the line width could be also precisely taken into account, reducing the associated uncertainty to a negligible level. Finally, at a relative uncertainty of $10^{-6}$, only the homogeneous broadening, which ranges between $10^{-2}$ and $10^{-4}$ $\Delta_D$ under our experimental conditions, contributes to the line width. The natural line width of the order of 1 Hz is negligible. The only parameter of importance is the pressure in the cell which will have to be carefully measured. The Gaussian, or Voigt, profile represents the absorption coefficient which coincides with the absorption signal only for an optically thin medium. For a good signal-to-noise ratio, we operate at higher pressures and the normalized line shape is $\exp(-\alpha L)$ where $L$ is the length of the absorption cell and $\alpha$ is the absorption coefficient. At low pressure, it is given by the Gaussian profile:

$$\alpha(\nu) = A n_0 \exp\left(-\frac{(\nu - \nu_0)^2}{\Delta(n_0)^2}\right)$$

where $n_0$ is the population density in the considered ground level, a quantity proportional to the pressure, and $A$ is a constant specific to our transition, $\nu$ is the laser frequency, $\nu_0$ is the resonance frequency and $\Delta(n_0)$ is the 1/e-fold half-width at of the Gaussian, which tends linearly to the Doppler width when the pressure (or $n_0$) tends to zero. As a consequence, the fit of the experimental line shape by an exponential of a Gaussian gives the two key parameters: the width of the Gaussian and the amplitude which is, in fact, in undetermined units, a measurement of the pressure. This way to calibrate the pressure is much more precise than by a pressure gauge and is particularly appropriate for an extrapolation to zero pressure. However, this parameter is related to the pressure of the active gas, while the width of the Gaussian depends on the total pressure including the impurities in the cell which must be eliminated as much as possible for a correct extrapolation to zero pressure.

**Results and discussion**

Fig. 2 displays a series of absorption spectra at pressures between 0.2 to 10 Pa. The absorption varies between 10% and 80% and at the highest pressures the line shape clearly differs from a Gaussian. The absolute pressure is measured with a Baratron gauge but this value is not used for the analysis of the results, as explained above. Typical signal-to-noise ratio (S/N) under the best conditions is $10^3$ for a time constant of 20 ms. A set of 2000 spectra were recorded over more than one month. Fig. 3 displays the linear regression which leads to the determination of the Doppler width by extrapolation to zero pressure. The relative calibration of the pressure axis is systematically performed by the amplitude of the signal as explained in the previous section. This uncertainty which is directly related to the S/N is negligible for the extrapolation. The resulting Doppler e-fold half-width is:

$$\Delta_D = 49.8831\,(47)\text{ MHz} \quad (9.5\times10^{-5})$$

This leads to the following determination of the Boltzmann constant:

$$k_B = 1.380\,65\,(26)\times10^{-23}\text{ J K}^{-1} \quad (1.9\times10^{-4})$$

The uncertainty takes into account that in the temperature. This value is in agreement with the CODATA value. Attempts to observe systematic effects due to the modulation index, the size or the shape of the laser beam, and the laser power were unsuccessful, as expected from the analysis of the previous section. For the fit, a slope is added to the exponential of the Gaussian to take into account a hypothetical background optical signal. A careful analysis of the data revealed that by suppressing the spectra which present a slight slope, the dispersion of the 500 remaining measurements was reduced, leading to the same result with the same uncertainty. On the other hand, regrouping data with a given slope gives a systematic shift. This is the only observation of a systematic effect, and is probably due to parasitic light reaching the detector. For further experiments, this observation offers a method to avoid this very tiny systematic effect by adjustment of the optical alignment.

**Conclusion and Perspectives**

This demonstration experiment of an optical method for the measurement of the Boltzmann constant which reaches an uncertainty of $2\times10^{-4}$ after a cumulative time of 61 hours is very promising since obvious improvements can be made in the near future. Three kinds of progress will be implemented: a temperature control with a stability better than $10^{-6}$; an absorption length 10 times larger in a multiple path cell will improve the extrapolation to zero pressure; changing the intensity stabilisation scheme will reduce the dead time between data points by a factor 10, thus reducing by almost the same factor the accumulation time of the spectra. These three straightforward directions of progress let us expect a gain of 2 orders of magnitude in accuracy comparable to that of the only measurement used in the CODATA. This method is very general and is a direct application of the first principles: the direct measurement of the thermal energy related to one degree of freedom of the system. In fact, the line profile reflects the velocity distribution along the laser beam axis for molecules in a given rovibrational level while the measured temperature is that of the whole sample. Thus, the obtention of a common value by using various transitions and temperatures leads to a partial check of the equipartition principle. Moreover, our method can be easily applied to different physical systems, different molecules and isotopes in a very large range of temperatures. One should also explore, especially at higher pressure, the influence of the adsorption energy on walls, the non-ideal character of the gas for a possible departure from a Maxwell-Boltzmann distribution. In addition, our experiment will open access to a detailed study of the Lamb-Dicke effect. Finally, the measure of the Doppler width in gases will give an universal way to measure the absolute thermodynamical temperature.

**Aknowledgment**
This work is supported by the Laboratoire National de Métrologie et d'Essais.

Figure caption

Fig 1: experimental set up. f=8 kHz, $f_1$=17 kHz, $f_2$=1,7 kHz, $f_3$=2,2 kHz. The sideband $\nu_{SB-}$ from the EOM is kept after filtering for recording the $NH_3$ absorption profile.

Fig 2: series of absorption spectra at different pressures

Fig 3: linear regression of the 1/e half-width of the absorption line which gives the extrapolated value at zero pressure, 49.883 1 (47) MHz.

Figure 1

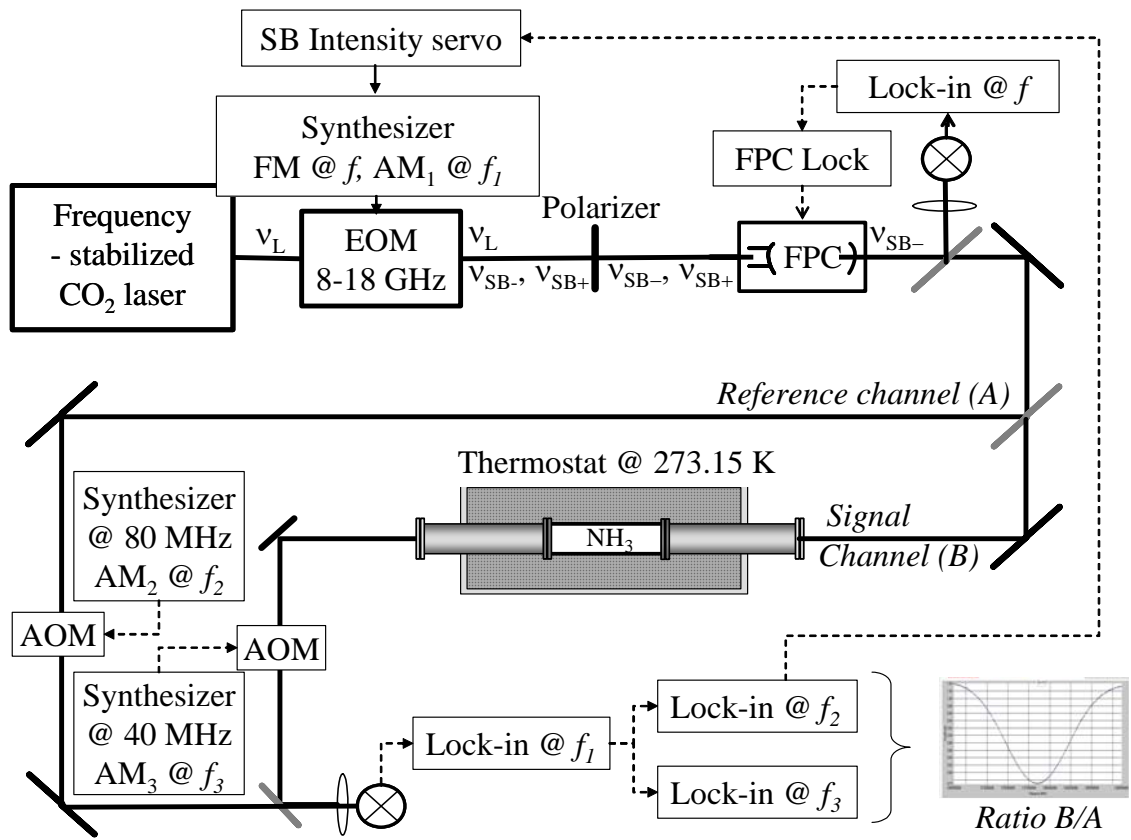

Figure 2 :

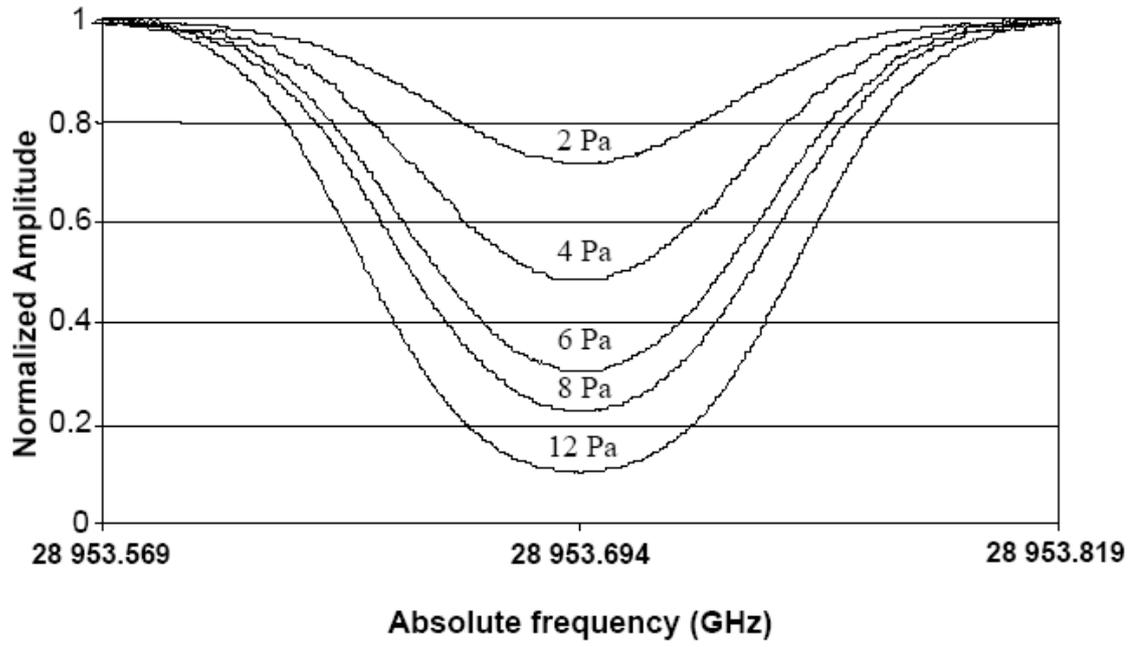

Figure 3

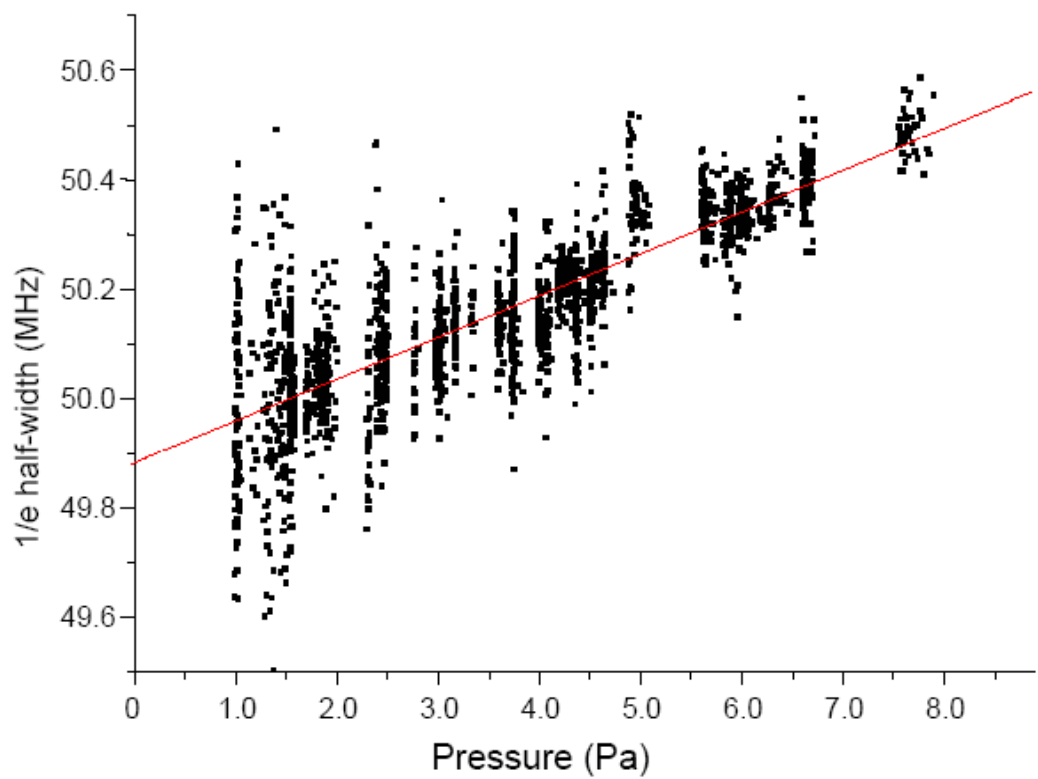